\documentstyle[nato]{crckapb}
\begin{opening}
\title{Signatures of Quantum Gravity in the Large-Scale Universe}
\author{L. P. Grishchuk}
\institute{Department of Physics and Astronomy, University of Wales 
Cardiff, Cardiff, CF2 3YB, United Kingdom\\
and\\
Sternberg Astronomical Institute,\\
Moscow University, 119899 Moscow V-234, Russia\\
\\
Based on a talk given at the 4-th Colloque Cosmologie,\\ 
Paris, June 1997}

\end{opening}

\begin{document}
The quantum gravity processes that have taken place in the very
early Universe are probably responsible for the observed large-scale 
cosmological perturbations. The comparison of the theory with the 
detected microwave background anisotropies favors the
conclusion that the very early Universe was not driven by a
scalar field with whichever scalar field potential. At the same
time, the observations allow us to conclude that there is a good
probability of a direct detection of the higher frequency relic 
gravitational waves with the help of the advanced laser 
interferometers.    

\section{Introduction. Fundamental Constants and Units}

The observable quantum gravity effects are usually considered as a 
subject of study for a very remote future. In any case, they are normally
expected to take place on microscopic, rather than macroscopic, scales.
It is remarkable that we are probably facing these effects right now, 
and on extremely large scales, in the form of very long-wavelength 
cosmological perturbations responsible for the observed~[1] 
large-angular-scale anisotropies in the cosmic microwave background (CMB) 
radiation. It is also likely that we will have direct access to these 
effects on smaller scales, and in the near future, by detecting 
the relic background of (squeezed) gravitational waves~[2] with the 
help of gravity-wave detectors currently under construction or in the phase 
of the design study (for a recent review, see~[3]). 

Since the mentioning of the subject of quantum gravity often instigates 
the feeling of disbelief and panic among the audience, we should first 
clarify our intentions. We are not going to quantize, create and destroy, 
universes. We will be discussing a conceptually simpler problem of 
quantizing cosmological perturbations in the Universe. 
This approach is quite analogous to, say, quantization of excitations 
in a sample of a condensed matter, or to the quantum field description 
of the squeezed light generation in quantum optics experiments~[4,5]. 

The cosmological perturbations (density perturbations, rotational
perturbations, gravitational waves) consist of a gravitational field
component and a matter component. The gravitational field and matter
variables are governed together by a system of coupled differential 
equations~[6]. In case of gravitational waves, only the gravitational
field component is present. An attempt of quantization of the matter 
part of cosmological perturbations without quantizing the associated 
gravitational field part would be as inconsistent as an attempt 
of quantization of the electric components of the electromagnetic field 
without quantizing the magnetic components. 

The central point of our effort is realization of the fact that even 
if there were no cosmological perturbations initially, they must have 
been generated later, under quite general conditions, as a result 
of parametric (superadiabatic) amplification of the zero-point quantum 
oscillations of the perturbation field~[7]. The source of amplification, 
which ``pumps'' energy into the zero-point (vacuum) fluctuations, 
is the strong variable gravitational field of the very early Universe.          

Although we are quantizing only the perturbed part of the gravitational 
field, we are still quantizing gravity. We expect our results to depend 
explicitly on all the three fundamental constants: 
$G$-representing gravity, $c$-representing relativity, and 
$\hbar$-representing quantum theory. 
Specifically, the $G$ and $c$ participate in the definition of 
the gravitational energy-momentum tensor, and the $\hbar$ enters 
because of our quantum normalization of the energy of each mode to
${1\over 2}\hbar\omega$, {\it i.e.,} to a ``half of the quantum in each 
mode''. These constants combine naturally in the 
Planck length $l_{Pl} =\sqrt{G\hbar/c^3}$ or in the Planck mass 
$m_{Pl}=\sqrt{\hbar c/G}$.
The vacuum fluctuations of the field in question can be visualized as
oscillations with a nonzero amplitude proportional to $\sqrt{\hbar}$.
Since everything in this formulation of the problem depend on 
the presence of the vacuum fluctuations, our results must vanish 
when $\hbar$ is sent to zero (in other words, 
no quantum theory --- no nonzero effects). 
It is important to know why the Planck constant $\hbar$ participates 
in our expressions, and it is helpful to write and keep track of all the 
fundamental constants. The unfortunate practice of setting outright
$G=c=\hbar =1$ is sometimes an indication that the author is not quite
sure why and which fundamental constants should be present, and whether,
say, the Planck constant should appear in the nominator or in the 
denominator of the final expression. So, for safety, so to say, 
all the fundamental constants are being set to 1. 
Similarly, the common practice of ``measuring'' some physical quantity 
(for instance, a scalar field) in units of the Planck mass is not at 
all an element of quantum physics, like the possibility of expressing 
the distance $L$ between the Earth and the Moon in units of the
Planck length, $L\approx 4\times 10^{43}l_{Pl}$, is not an indication 
of quantum nature of that distance.         	  

In our study, the fundamental constants $G$, $c$, $\hbar$ are present 
because we are quantizing the (perturbed part of) gravitational field 
and the associated matter fields (unless we are considering specifically 
gravitational waves in which case only gravity is being quantized). 
However, the $G$, $c$, $\hbar$ can also participate in the final 
expressions if a discussed problem does not belong to the realm 
of quantum gravity, but involves gravity plus quantum theory. 
A beautiful example is provided by the expression for the maximal 
masses and radii of cold white dwarfs and neutron stars. 
These objects owe their existence to the interplay between gravity 
and quantum mechanics. It is the pressure of the degenerate Fermi 
gas which through the Pauli exclusion principle enables the cold white
dwarfs and neutron stars to resist gravitational collapse~[8]. 
If there were no quantum mechanics $(\hbar =0)$, these objects 
would not exist. This statement in words is of course supported 
by formulae. The maximal masses of cold white dwarfs (the Chandrasekhar
mass) and neutron stars are approximately equal to 
\[
  M_{\rm max} \approx m_{Pl} \left( {m_{Pl}\over m_B}\right)^2
              \approx 1.5M_{\odot}
\]
where $m_B$ is the mass of a baryon. The equilibrium radius of a white
dwarf with the mass $M_{\rm max}$ is approximately equal to 
\[
   R \approx {\hbar \over m_e c}\left( {m_{Pl}\over m_B} \right)
     \approx 5\times 10^8~{\rm cm}
\]
where $m_e$ is the mass of an electron. These formulae explicitly
demonstrate that $M_{\rm max}$ and $R$ would vanish if the Planck 
constant were sent to zero. It is important to note that the sign 
of approximate equality in these formulae means the approximate equality. 
It is not a sign of proportionality, which would allow many other 
dimensional quantities to be present, and it is not a sign of an estimate
based on the dimensional analysis only, which would allow the numerical 
coefficient to be, say, 50 times larger or 50 times smaller than 1. 
A beautiful physics is usually expressed in simple and beautiful formulae: 
the final result depends essentially on the fundamental constants 
and a couple of parameters, such as $m_B$ and $m_e$, characterizing 
the system. As we will see below, our predictions for quantum-mechanically 
generated cosmological perturbations in simple cosmological models 
do also depend essentially on the fundamental constants, combined in 
$l_{Pl}$, and a couple of parameters characterizing the model.

The inspection of positions and powers in which the fundamental constants 
enter the claimed result may serve as a first test of whether we are 
prepared to believe the result. For instance, when you read that 
``inflation generates'' density perturbations because it 
``stretches the vacuum fluctuations beyond the horizon'', 
plus accompanying words about something quantum becoming something 
classical ``upon the horizon crossing'', this explanation may raise 
your doubts even before any further reading. If everything is so simple, 
why do we consider vacuum fluctuations of the ``inflaton'', and not 
the vacuum fluctuations of the electromagnetic field which we definitely 
know to exist? Since the vacuum fluctuations of the electromagnetic 
field are being ``stretch beyond the horizon'' in exactly the same manner, 
we should apparently be able to generate enormous amount of photons as well. 
However, the result for photons is known:  strict zero, independently 
on whether the expansion is inflationary or not. 
This example simply shows that it is the dynamical property of the 
system, that is, whether and how the vacuum fluctuations are coupled 
to the pumping field, and not the universal kinematical property of 
``stretching'' the wavelength ``beyond the horizon'', which is really 
important. As a first test of the inflationary predictions you may 
wish to check whether and how the Planck constant enters the claimed result. 
Surprisingly, you will not find the Planck constant, or when it is present 
in the form of the Planck mass, it appears in the denominator, 
and not in the nominator, of the final expression. 
Taken for the face value, this position of the Planck constant would 
suggest that the effect, owing its existence to the quantum fluctuations, 
goes to infinity when the Planck constant is sent to zero. 
Of course, the correct position of the fundamental constants is not 
a guarantee that the result is correct numerically. 
And the lack, or a strange position, of the fundamental constants 
in the final expression is not an indication that the result 
is necessarily wrong. The author could have set $G=c=\hbar=1$ everywhere, 
or could have done this in one part of the participating equations but 
forgot to do the same in the other. However, an unclear physical 
explanation of the expected effect and a chaotic placing of the 
fundamental constants in the resulting expression does raise a suspicion 
that the numerical result may also be wrong. We will return to this point 
later. 

\section{Quantum Normalization}

In order to quantize cosmological perturbations in a traditional
field-theoretical manner, it is convenient to use the entire general
relativity formulated in a traditional field-theoretical manner 
(see~[9] and the early papers quoted there).
The field-theoretical formulation of the Einstein gravity operates with
the gravitational field potentials $h_{\mu\nu}$ (dimensionless, 
symmetric, second rank tensor) and matter field variables. 
Both sets of variables are specified in the Minkowski space-time
\begin{equation}
    d\sigma^2 = \eta_{\mu\nu}~dx^\mu~dx^\nu
  = c^2dt^2 - {dx^1}^2 - {dx^2}^2 -{dx^3}^2 \quad .
\end{equation}
The total Lagrangian $L$ of the theory consists of the 
gravitational part $L^g$, which contains only gravitational 
variables, and the matter part $L^m$, which contains matter 
variables and gravitational variables.  The concrete form of
$L^g$ and $L^m$, which we do not write here,
distinguishes the Einstein gravity from many possible alternative theories. 
The exact Einstein equations follow from the variational principle and have
the form
\begin{equation}
  {{h^{\mu\nu}}_{,\alpha}}^{,\alpha}
 -{{h^{\mu\alpha}}_{,\alpha}}^{,\nu}
 -{{h^{\nu\alpha}}_{,\alpha}}^{,\mu}
 +\eta^{\mu\nu}~{{h^{\alpha\beta}}_{,\alpha ,\beta}}
 = 2\kappa(t^{\mu\nu} + \tau^{\mu\nu})
\end{equation}
where $t^{\mu\nu}$ is the energy-momentum tensor of the gravitational field, 
$\tau^{\mu\nu}$ is the energy-momentum of the matter fields including
their interaction with gravity, and $\kappa =8\pi G/c^4 $. 
The energy-momentum
tensors $t^{\mu\nu}$ and $\tau^{\mu\nu}$ are defined as variational 
derivatives of $L^g$ and $L^m$, correspondingly, 
with respect to the Minkowski metric.

If you want to arrive at the geometrical formulation of the general
relativity, you are welcome to introduce new functions $g_{\mu\nu}$ 
according to the rule                            
\begin{equation}
  \sqrt{-g}~ g^{\mu\nu} = \eta^{\mu\nu} + h^{\mu\nu}
\end{equation}
and to interpret these new functions as the metric tensor of the 
curved space-time
\begin{equation}
   ds^2 = g_{\mu\nu}~dx^\mu~dx^\nu \quad .
\end{equation}
The substitution (3) translates the Lagrangian $L^g$ 
into the Hilbert-Einstein Lagrangian (up to a total derivative) 
and the field equations (2) into the usual geometrical Einstein equations. 
How the universal gravitational field $h_{\mu\nu}$ affects equations 
of motion of matter fields, and how and why, on the grounds of physical 
measurements, we conclude that we ``live'' in the curved space-time (4), 
is a subject of a separate discussion~[9].        

In addition to the coordinate freedom, {\it i.e.,} the freedom of choosing 
arbitrary (curvilinear) coordinates in the Minkowski space-time (1), 
the field-theoretical formulation of gravity allows also the gauge 
freedom, {\it i.e.,} special re-definitions of gravitational 
and matter variables which do not change the field equations (2). 
We may use this gauge freedom, and we will sometimes do that, 
in order to impose the conditions
\begin{equation}
  {h^{\mu\nu}}_{,\nu} = 0
\end{equation}
and to further simplify equations (2).

From the field-theoretical perspective, cosmological solutions are
nothing more than specific distributions of gravitational 
and matter fields in the Minkowski space-time (1). 
Concretely, the homogeneous isotropic FLRW models
\begin{equation}
          ds^2 = c^2 dt^2 - a^2(t)
  \bigl( {dx^1}^2 + {dx^2}^2 + {dx^3}^2 \bigr)
\end{equation}
are described by the gravitational potentials 
\begin{equation}
   h_{00} = a^3(t) -1~~~,\qquad
   h_{11} = h_{22} = h_{33} = 1-a(t)\quad .
\end{equation}
Of course, while operating with the Minkowski metric (1), we do not 
mean that the measurable distance between two objects located at $x$ 
and $x+dx$ is equal to $dx$. In fact, the gravitational field (7) 
makes the measurable distance equal to $a(t)dx$. 
This distance would have been equal to $dx$ if there were no gravitational 
field (7), {\it i.e.,} for $a(t)=1$.  But this is a different issue. 
At the moment, we are interested in other sides of the theory.

In the geometrical approach, the gravitational part of cosmological 
perturbations is conveniently written as
\begin{equation}
  ds^2 = c^2dt^2 -a^2(t)(\delta_{ij}+h_{ij})dx^i~dx^j \quad .
\end{equation}
In the field-theoretical approach, we write
\begin{equation}
  h_{\mu\nu} 
= h_{\mu\nu}^{(0)} + h_{\mu\nu}^{(1)} \quad ,
\end{equation}
where $h_{\mu\nu}^{(0)}$ are given by Eq.~(7). The functions 
$h_{ij}(x^0,x^1,x^2,x^3)$
participating in Eq.~(8) and the functions
$h_{\mu\nu}^{(1)}(x^0,x^1,x^2,x^3)$
participating in Eq.~(9) are totally different objects. 
They have differing transformation properties and differing meanings. 
But, as functions of 
$(x^0,x^1,x^2,x^3)$, 
they are simply related in the linear approximation that we consider.  

In case of gravitational waves, we can write a spatial Fourier component
of the field $h_{\mu\nu}^{(1)}$ for a given wave-vector ${\bf n}$ 
in the form:
\begin{equation}
   h_{ij}^{(1)} = \sum_{s=1}^2 {\stackrel{s}{p}}_{ij}
\left[ {\stackrel{s}{\mu}}_{\bf n}(x^0)~e^{i{\bf n}\cdot{\bf x}}
      +{\stackrel{s}{\mu^\ast}}_{\bf n}(x^0)~e^{-i{\bf n}\cdot{\bf x}}
\right]~~.
\end{equation}
When writing Eq.~(10), the gauge conditions (5) and the remaining gauge 
freedom preserving (5) have been used in order to eliminate         
$h_{oo}^{(1)}$ and $h_{oi}^{(1)}$, and to satisfy the requirements
\begin{equation}
 {\stackrel{s}{p}}_{ij}n^j =0~~,\qquad
 {\stackrel{s}{p}}_{ij}\delta^{ij} =0 \quad ,
\end{equation}
where 
${\stackrel{s}{p}}_{ij}$ $(s=1,2)$
are constant polarization tensors obeying the conditions
\begin{equation}
  {\stackrel{s}{p}}_{ij}~
  {\stackrel{s'}{p^{ij}}} = 2\delta_{ss'}~~~, \qquad
  {\stackrel{s}{p}}_{ij}({\bf n})
 ={\stackrel{s}{p}}_{ij}(-{\bf n})~~~.
\end{equation}
In case of gravitational waves, the functions 
$h_{ij}$ and $h_{ij}^{(1)}$ 
are related as\\
$h_{ij}^{(1)}=a(t)h_{ij}$, 
that is, Eq.~(10) can also be written as
\begin{equation}
  h_{ij} = {1\over a} \sum_{s=1}^2 
      {\stackrel{s}{p}}_{ij}
\Big[ {\stackrel{s}{\mu}}_{\bf n} (x^0)~e^{i{\bf n}\cdot{\bf x}}
     +{\stackrel{s}{\mu}}_{\bf n}^\ast (x^0)~e^{-i{\bf n}\cdot{\bf x}}
\Big] ~~.
\end{equation}

After the introduction of a new time parameter $\eta$ according to 
the relationship $d\eta =cdt/a(t)$, and denoting $d/d\eta$ by a prime,
the linear approximation to Eq.~(2) produces the familiar equation~[7]
\begin{equation}
      \mu^{\prime\prime} + \mu
\left[ n^2-{a^{\prime\prime}\over a} \right] = 0
\end{equation}
for every mode ${\bf n}$ and for each polarization state $s$. 
This equation has two linearly independent solutions with two arbitrary 
numerical coefficients in front of them. In the high-frequency regime,
that is, when the interaction of the waves with the pump field (7) 
is negligibly small 
$n^2 \gg \big| {a^{\prime\prime}\over a}\big|$, the general solution 
to Eq.~(14) has the form
\begin{equation}
  \mu(\eta ) \approx A_1~e^{-in\eta} + A_2~e^{in\eta}\quad .
\end{equation}

Classically, the constants $A_1$, $A_2$ are arbitrary numbers 
determined by the initial conditions. If there is no classical waves, 
$|A_1|=|A_2| =0$, there is nothing to interact with the pump field, 
and there is nothing to be amplified. Quantum-mechanically, 
we expect the initial amplitude of the waves to be nonzero
and to be determined by the zero-point quantum oscillations. 
We want to assign to these words a concrete meaning. 
We want to determine the numerical coefficient $C$ in the general 
expression for the field operator
$h_{ij}^{(1)} (\eta ,{\bf x})$: 
\begin{equation}
       h_{ij}^{(1)}(\eta ,{\bf x}) = {C\over (2\pi)^{3/2}}
       \int_{-\infty}^\infty d^3{\bf n} \sum_{s=1}^2
       {\stackrel{s}{p}}_{ij}({\bf n}) {1\over\sqrt{2n}}
\Big[ {\stackrel{s}{c}}_{\bf n}(\eta) e^{i{\bf n}\cdot{\bf x}}
     +{\stackrel{s}{c}}_{\bf n}^\dagger (\eta) e^{-i{\bf n}\cdot{\bf x}}
\Big] ,
\end{equation}
where the creation and annihilation operators satisfy the commutation  
rules 
$[{\stackrel{s}{c}}_{\bf n}(\eta ),~
{\stackrel{s'}{c}}_{\bf m}^\dagger (\eta )]
=\delta^3 ({\bf n}-{\bf m})\delta_{ss'}$, 
and obey the Heisenberg equations of motion with the corresponding 
Hamiltonian. 

We need to calculate the contribution of gravitational waves 
to the gravitational energy-momentum tensor $t^{\mu\nu}$ 
(whose exact form is known~[9]) in quadratic approximation. 
Since, at the moment, we are interested only in the normalizing 
coefficient $C$, we can consider the high-frequency waves. 
Formally, we simply set $a(t)=1$ and derive the quadratic approximation 
for $t^{\mu\nu}$ using the fact that the functions 
$h_{\mu\nu}^{(1)}$
satisfy the gauge conditions mentioned  above, and that the frequency
$w_n$ of the wave with the wave number $n$ is
$w_n=cn$.
Then, the quadratic expression for $t^{\mu\nu}$ reduces to
\begin{equation}
  t_{\mu\nu} = {1\over 4\kappa} 
    {h^{ij(1)} }_{,\mu} ~h_{ij~,\nu}^{(1)} \quad .
\end{equation}

The quantum-mechanical operator for $t_{\mu\nu}$ is constructed by
inserting Eq.~(16) into Eq.~(17). The initial vacuum state for each 
mode of the field is defined by the requirement 
${\stackrel{s}{c}}_{\bf n} (0)|0\rangle = 0$,
where 
${\stackrel{s}{c}}_{\bf n}(0)$
and 
${\stackrel{s}{c}}_{\bf n}^\dagger (0)$
are  the initial values of the annihilation and creation operators,
that is, their values taken long before any significant interaction 
with the pump field has occurred. The constant $C$ is determined by 
the requirement that the energy of the field is equal initially to 
a ``half of the quantum per mode'', that is 
\[
   \langle 0|\int_{-\infty}^\infty t_{oo} d^3{\bf x} |0\rangle
 = {1\over 2}\hbar \int_{-\infty}^\infty d^3{\bf n}~w_n
   \sum_{s=1}^2 \langle 0|
   {\stackrel{s}{c}}_{\bf n}(0)
   {\stackrel{s}{c}}_{\bf n}^\dagger (0)
 + {\stackrel{s}{c}}_{\bf n}^\dagger (0)
   {\stackrel{s}{c}}_{\bf n}(0)|0\rangle ~.
\]
From this requirement we derive
$C=\sqrt{16\pi}~l_{Pl}$. 
Everything in our problem is well defined now. 

In case of density perturbations, the gravitational field part and
the matter part of the perturbations are two sides of the same physical
entity. For simple models of matter, such as scalar fields and perfect
fluids, the full set of perturbed Einstein equations can be reduced~[10] 
to a single second order differential equation similar to Eq.~(14).
All the functions describing the perturbations can then be found from 
solutions to this equation via simple algebraic and 
differentiation-integration relationships, to which the rest of equations 
amounts. Concretely, a spatial Fourier component of the metric 
perturbations $h_{ij}$, defined by Eq.~(8), can be written in the form 
similar to Eq.~(13):  
\begin{equation}
  h_{ij} = \sum_{s=1}^2 {\stackrel{s}{p}}_{ij}
\left[ {\stackrel{s}{h}}_{\bf n}(\eta)~e^{i{\bf n}\cdot{\bf x}}
      +{\stackrel{s}{h}}_{\bf n}^\ast (\eta)~e^{-i{\bf n}\cdot{\bf x}}
\right]
\end{equation}
where the polarization tensors 
${\stackrel{s}{p}}_{ij}$ 
obeying the conditions (12) are now
\[
    {\stackrel{1}{p}}_{ij} 
  = \sqrt{{2\over 3}}~\delta_{ij} ~~,\qquad
    {\stackrel{2}{p}}_{ij} 
  = -\sqrt{3} ~{n_in_j\over n^2} + {1\over \sqrt{3}}~\delta_{ij} ~~~.
\]
The component 
${\stackrel{2}{h}}_{\bf n} (\eta )$ 
of the metric perturbations is not an independent function, 
and the perturbed Einstein equations allow us to express this function 
through the scalar component 
${\stackrel{1}{h}}_{\bf n} (\eta )$ 
of the metric perturbations. We denote the scalar component 
${\stackrel{1}{h}}_{\bf n} (\eta )\equiv h_{\bf n}(\eta )$ 
and omit the mode label ${\bf n}$.

It is convenient to introduce a new function $\overline{\mu}(\eta )$ 
through the definition
\begin{equation}
       h + {a\over a^\prime\gamma} h^\prime 
\equiv {\overline{\mu} \over a} ~~~,
\end{equation}
where
\[
  \gamma \equiv 1+ \left( {a\over a^\prime} \right)^\prime
         \equiv -{\dot{H} \over H^2}~~~,
\]
and $H$ is the Hubble parameter, 
$H\equiv \dot{a} / a$.
The function $h(\eta )$ changes under the small coordinate transformations 
respecting the form of Eq.~(8), but the function 
$h+(a/a^\prime\gamma)h^\prime$
and, hence, the function 
$\overline{\mu} / a$ do not change. 
In this sense, the function 
$\overline{\mu} / a$
is the coordinate-invariant core of the metric component 
$h(\eta)$. 

If the matter is a (minimally coupled) scalar field with arbitrary 
scalar field potential, the single second-order differential equation 
mentioned above is the equation for the function $\mu (\eta )$, 
where $\mu\equiv\sqrt{\gamma}~\overline{\mu}$:
\begin{equation}
       \mu^{\prime\prime} + \mu
\left[ n^2 -
      {(a\sqrt\gamma )^{\prime\prime} \over a\sqrt\gamma}
\right] = 0 \quad .
\end{equation}
This equation defines the entire dynamical content of the problem 
and is the only one equation which needs to be solved~[10]. 
The components 
${\stackrel{1}{h}}_{\bf n}(\eta )$
and
${\stackrel{2}{h}}_{\bf n}(\eta )$
of the perturbed metric, as well as the scalar field perturbation, 
are all expressible through the solutions of this equation. 
The function 
$\overline{\mu} / a$ 
is reminiscent of  the 
electrodynamic potential allowing to express the electric and magnetic 
components of the electromagnetic field through the solutions for the
potential. It follows from Eq.~(20) that in the long-wavelength regime
the dominant solution is $\mu\sim a\sqrt\gamma$. 
That is, the function 
$\overline{\mu} / a$ 
is approximately constant, in full analogy with the
approximate constancy of the metric amplitude $\mu / a$
for gravitational waves, Eqs.~(13), (14). 

After having found a single independent degree of freedom, 
completely characterizing our coupled system of metric and scalar field 
perturbations, we can proceed with the quantization, in the manner similar 
to that of gravitational waves. It is important to note that 
if the pump field (7) is such (or, in other words, if the scale factor
$a(\eta )$ is such) that the function $\gamma(\eta )$ becomes 
a constant, Eq.~(20) reduces to exactly the same form as Eq.~(14) 
for gravitational waves. In any case, the quantum-mechanical
operator for 
$h_{ij}(\eta ,{\bf x})$
can be written in the form similar to Eq.~(16): 
\[
        h_{ij}(\eta ,{\bf x})
      = {C \over (2\pi)^{3/2}} ~{1\over a(\eta )}
        \int_{-\infty}^\infty d^3{\bf n} \sum_{s=1}^2
        {\stackrel{s}{p}}_{ij}({\bf n}) {1\over \sqrt{2n}}
\Big[ {\stackrel{s}{c}}_{\bf n}(\eta )e^{i{\bf n}\cdot{\bf x}}
      +{\stackrel{s}{c}}_{\bf n}^\dagger 
       (\eta )e^{-i{\bf n}\cdot{\bf x}}
\Big].
\]
\begin{equation}
\end{equation}
For the density perturbations, however, the operator 
${\stackrel{2}{c}}_{\bf n}(\eta )$
is not truly independent and is expressible through 
${\stackrel{1}{c}}_{\bf n}(\eta )$.
In its turn, the operator 
${\stackrel{1}{c}}_{\bf n}(\eta )$
is expressible with the help of Eq.~(19)  through the operator
$d_{\bf n}(\eta )$
for the single degree of freedom, mentioned above, 
\begin{equation}
  {\stackrel{1}{c}}_{\bf n}(\eta )
= {a^\prime \over a} \int\gamma(\eta )d_{\bf n}(\eta )~d\eta ~~~,
\end{equation}
which is subjected to quantization:
\begin{equation}
   \big[ d_{\bf n}(\eta ),d_{\bf m}^\dagger (\eta )\big]
 = \delta^3({\bf n}-{\bf m}) ~~~,\quad
   d_{\bf n} (0)|0\rangle = 0 ~~~~.
\end{equation}
The operator $d_{\bf n}(\eta )$ is the operator-valued version 
of the function $\overline{\mu}_{\bf n}(\eta )$.
The normalization constant $C$ in Eq.~(21) has been found~[10] 
to be equal\\
$C=\sqrt{24\pi}~l_{Pl}$.
Thus, the quantum normalizing coefficient $C$ for the metric 
perturbations in the case of scalar field density perturbations 
is practically the same as in the case of gravitational waves.  

In this presentation, we are working with the spatially-flat
models (6) but the quantum-mechanical generating mechanism 
is of course independent of this assumption.

\section{Metric Perturbations at the Matter-Dominated Stage}

The (operator-valued) metric perturbations $h_{ij}$, Eq.~(8), 
at the matter-dominated stage are essentially all we need to know. 
In case of density perturbations and gravitational waves, 
a synchronous coordinate system (8) can, in addition, be made 
the comoving coordinate system, what we assume to be done. 
The functions $h_{ij}$ enter the calculation of the microwave 
background anisotropies, and these functions allow us also to find, 
due to the perturbed Einstein equations, other quantities, 
such as the perturbation 
$\delta\rho /\rho$ 
in the matter mass density. In the Schr\"odinger picture, 
the quantum-mechanical evolution of the initial vacuum state 
$|0_{\bf n}\rangle |0_{-{\bf n}}\rangle$
results in the appearance of the two-mode squeezed vacuum quantum states. 
With them are associated the specific statistical properties 
of the perturbation field itself and the microwave  background 
anisotropies caused by the field~[10]. However, we will first
evaluate the characteristic amplitude of the metric perturbations. 
The characteristic amplitude $h(n)$ is a dimensionless quantity defined 
as the dispersion (square root of variance), or in other words, 
the root mean squared amplitude of the field, per logarithmic 
frequency interval. We use here the Heisenberg picture, in which the
initial vacuum state $|0\rangle$ remains fixed, while the dynamical 
evolution is carried by the time-dependent field operators. 
In general, the $h(n)$ is defined through the equality
\[
    \langle 0|h_{ij}h^{ij}|0\rangle 
 = \int_0^\infty h^2(n) {dn\over n}\quad .
\]
For gravitational waves, it is convenient to describe the characteristic 
amplitude $h(\nu)$ as a function of the frequency $\nu$ measured in Hz. 
For density perturbations, it is more appropriate to retain the
original dependence $h(n)$ on the corresponding wave-number 
(spatial scale) $n$, since density perturbations at the 
matter-dominated stage do not oscillate as functions of time.   

The present matter-dominated stage of cosmological evolution
$a(\eta )\sim\eta^2$
was preceded by the radiation-dominated stage 
$a(\eta )\sim\eta$.
How the Universe behaved at still earlier times, that is, well before
the era of primordial nucleosynthesis, is not known.  Following
S.~Weinberg, this initial stage of evolution is called the very early
Universe. Usually, we describe the unknown evolution of the very early
Universe by simple power-law scale factors. These are quite representative
models, because, if necessary, a more complicated evolution can be
approximated by a series of power-law pieces. For purposes of
numerical evaluations, it is convenient to regard 
$\eta,~x^1,~x^2,~x^3$ as dimensionless coordinates and to make 
the scale factor $a(\eta )$ carrying the dimensionality of length.

A large class of expanding models is described by the scale factors
\begin{equation}
  a(\eta ) = l_o|\eta |^{1+\beta} \quad ,
\end{equation}
where $\eta$ time grows from $-\infty$ and the constant parameter 
$\beta$ is $\beta < -1$ at the initial stage of evolution. 
The constant $l_o$ has the dimensionality of length. 
With this scale factor the function $\gamma (\eta )$ becomes a constant:
\begin{equation}
   \gamma = {2+\beta \over 1+\beta} \quad .
\end{equation}
The Einstein equations require the effective equation of state for
the matter driving the evolution (24) to be in the form
\begin{equation}
    p = {1-\beta \over 3(1+\beta )} \epsilon \quad .
\end{equation}
The case $\beta =-2$ corresponds to $p=-\epsilon$ and to the
de Sitter scale factor $a(\eta )=l_o|\eta |^{-1}$.
	
Having specialized the cosmological evolution $a(\eta )$ and
knowing the dynamical equations for cosmological perturbations, as
well as their initial values dictated by the quantum normalization, 
we can find the characteristic amplitudes $h(\nu )$ and             
$h(n)$ at any given time $\eta$. 
When calculating the $h(\nu )$ and $h(n)$,
the important property of solutions to Eqs.~(14), (20) is being used:
in the long-wavelength regime, the function 
$\mu / a$ satisfying Eq.~(14), as well as the function 
$\overline{\mu} / a$ 
satisfying Eq.~(20), stay practically constant. 
For definiteness, we will give estimates for the present time, 
but it is clear how the functions $h(\nu )$ and $h(n)$ 
scale at earlier times. We will also ignore the oscillations in
the functions $h(\nu )$ and $h(n)$ which always take place for 
high-frequency and, correspondingly, short-wavelength parts of these
spectra as a result of squeezing and standing-wave pattern of the
perturbations.

In case of gravitational waves, the results are as follows.  
For $\nu\leq \nu_H$, where $\nu_H\approx 10^{18}$~Hz 
is the present-day Hubble frequency,
\begin{equation}
  h(\nu ) \approx {\l_{Pl}\over l_o}
  \left( {\nu \over \nu_H}\right)^{2+\beta} \quad .
\end{equation}
Specifically for the (de Sitter) model $\beta = -2$, the function (27)       
is independent of frequency $\nu$, {\it i.e.,} we obtain a ``flat'' 
part of the spectrum. The constant $l_o$ in this model gives 
the value of the (constant) Hubble parameter at the de Sitter stage,
$H= c / l_o $, and determines the (constant) curvature of the 
de Sitter space-time. 

In the interval 
$\nu_H\leq \nu\leq \nu_m$, where the frequency $\nu_m$ 
is defined by the time (energy density) of transition from the 
radiation-dominated to the matter-dominated stage, 
$\nu_m\approx 10^{-16}$~Hz,
\begin{equation}
       h(\nu ) \approx {l_{Pl}\over l_o}
\left( {\nu\over\nu_H}\right)^\beta \quad .
\end{equation}

In the interval 
$\nu_m\leq\nu\leq\nu_c$, 
where the frequency $\nu_c$ is defined by the time (energy density) 
of transition from the initial stage of expansion (24) to the 
radiation-dominated stage and above which the spectrum falls sharply
since the waves with frequencies higher than $\nu_c$ are not affected 
by the amplification process, $\nu_c\approx 10^8$~Hz in the currently 
discussed models,
\begin{equation}
        h(\nu ) \approx {l_{Pl} \over l_o}
\left( {\nu_m \over\nu_H}\right)^\beta 
\left( {\nu \over\nu_m}\right)^{1+\beta } \quad .
\end{equation}

As we see, the scale factors (24) which are power-law dependent 
on $\eta$ time generate spectra which are power-law dependent on 
frequency $\nu$~[7]. The final result depends on the fundamental
constants, combined in $l_{Pl}$, and a couple of cosmological parameters,
such as $l_o$ and $\beta$, characterizing the model. The sending of
the Planck constant to zero would of course eliminate the entire
expression for $h(\nu )$.          

In case of density perturbations, the characteristic amplitude $h(n)$
for wavelengths longer than, and comparable with, the Hubble
radius is given by formula similar to Eq.~(27):
\begin{equation}
  h(n) \approx {l_{Pl} \over l_o}
\left( {n\over n_H}\right)^{2+\beta} \quad ,
\end{equation}
where the wavenumber $n_H$ corresponds to the wavelength
$\lambda_H = 2\pi a / n_H$ equal to the Hubble radius. 
The present-day $\lambda_H$ is $\lambda_H\approx 2\times 10^{28}$~cm.
Specifically for the $\beta =-2$ model, the function (30) is independent
of $n$, that is, we have a ``flat'' (or Harrison-Zeldovich-Peebles) 
part of the spectrum. (The slope of the spectrum for density perturbations, 
but not the amplitude (see below), is a correct part of conclusions in the 
influential early papers~[11].) 

The similarity of results (27) and (30) is a consequence of the 
following facts~[10]: 
(i)~the basic dynamical equations (14) and (20) 
strictly coincide for the scale factors (24), 
(ii)~the quantum normalizing coefficient $C$ of metric perturbations 
is practically the same quantity for density perturbations and 
for gravitational waves, and (iii)~the evolution of the long-wavelength 
metric perturbations associated with density perturbations or with 
gravitational waves is practically the same (namely, no evolution 
at all, the metric amplitudes stay almost constant)
all the way from the initial stage and up to the matter-dominated stage
({\it i.e.,} from the time when a given wave ``leaves'' the Hubble 
radius and up to the time when it ``enters'' the Hubble radius again). 
The numerical value of the characteristic amplitude at scales comparable 
with the Hubble radius is approximately the same number, both 
for gravitational waves and for density perturbations. Namely,
\begin{equation}
    h(\nu_H) \approx {\l_{Pl} \over l_o}
\end{equation}
for gravitational waves, and
\begin{equation}
  h(n_H) \approx {\l_{Pl}\over l_o}
\end{equation}
for density perturbations. This is true independently on how close was
the parameter $\beta$ to the de Sitter value $\beta =-2$ at the time 
when the wavelengths of our interest were ``leaving'' the Hubble radius 
at the initial stage of expansion. This conclusion is in a severe conflict
with the conclusion of inflationary literature.

The conclusion of the inflationary literature is that, at scales
comparable with the present-day Hubble radius, the amplitude of the
perturbation in the matter density $ \delta\rho / \rho$ is equal to 
\begin{equation}
  {\delta\rho \over \rho}\Bigg|_H \approx {H^2\over \dot{\phi}_o}~~~,
\end{equation}
where $\phi_o$ is the unperturbed scalar field, and the right hand side of
Eq.~(33) is supposed to be evaluated at the time when the wavelength of 
our interest was ``leaving'' the Hubble radius at the inflationary stage.
In the most recent literature, one can find the often quoted formula, 
which explicitly includes the Planck mass:
\begin{equation}
  \delta^2_H(k) = { 1\over 75\pi^2~m_{Pl}^6}~
  {V^3\over V^{\prime 2}}
\end{equation}
where $V$ is the scalar field potential and $V^\prime =dV/d\phi$.

Leaving aside the lack or strange position of the fundamental 
constants in these formulas, let us first explore them in the vicinity
of the most favorite (de Sitter) inflationary model. 
Since the denominator in formulas (33), (34) goes to zero in the limit
of the de Sitter expansion at the initial stage,
$\dot{\phi}_o\rightarrow 0$, $V' \rightarrow 0$,
these formulas predict arbitrarily large amplitudes for
$\delta\rho /\rho $ today.
The same conclusion follows for the characteristic metric amplitude,
because
$h(n)\approx \delta\rho / \rho$ when the density perturbations 
``enter'' the Hubble radius at the matter-dominated stage. 
By analyzing the line of the inflationary argumentation and reinstating 
the fundamental constants, one can find that, instead of Eqs.~(30) and (32), 
the inflationary prediction amounts to
\begin{equation}
         h(n)_{\rm infl}
\approx {1\over \sqrt\gamma}~{l_{Pl}\over l_o}
\left(  {n\over n_H}\right)^{-\gamma}
\end{equation}
for small $\gamma$. That is, formula (35) predicts larger and larger 
amplitudes $h(n)$ for models with smaller and smaller values of $\gamma$ 
at the initial 
stage of expansion ({\it i.e.,} for $2+\beta\rightarrow 0$, see Eq.~(25)). 
Since the long-wavelength metric perturbations are almost constant in time, 
the divergent as $1/\sqrt\gamma$ behaviour (35) must be postulated 
as the initial condition (at the first ``horizon-crossing'').

The inflationary claim about density perturbations leads to an incorrect 
conclusion regarding the possible relative contributions of the 
quantum-mechanically generated gravitational waves and density 
perturbations to the observed large-angular-scale anisotropies of CMB.
It is known~[12,13,14] that a numerical value of $h(\nu )$ at 
$\nu\approx \nu_H$ or $h(n)$ at $n\approx n_H$ 
translates into a numerical estimate for the quadrupole anisotropy 
$\delta T/T$.  According to Eqs.~(31), (32) the
contributions of gravitational waves and density perturbations are of 
the same order of magnitude. More accurate evaluation gives some numerical
preference to gravitational waves~[10]. In contrast, the inflationary 
claim about density perturbations transforms into the so-called consistency 
relations which are usually formulated as the statement that the 
ratio $T/S$ of the gravity-wave contribution $(T)$ to the 
contribution of density perturbations $(S)$ goes to zero, 
$T/S\rightarrow 0$, when the spectral index of density 
perturbations approaches the most favorite (Harrison-Zeldovich-Peebles) 
value, that is, in the limit 
$2+\beta\rightarrow 0$, $\gamma\rightarrow 0$.
Obviously, there is nothing wrong with the (small) finite contribution 
of gravitational waves in this limit: 
$\delta T/T\approx l_{Pl}/l_o$. 
This result for gravitational waves has never been a matter of dispute.
It is the claimed divergent contribution of density perturbations which
is responsible for $T/S\rightarrow 0$. 
The unacceptable situation with the inflationary prediction for density 
perturbations does not become less disturbing if it is formulated 
as $T/S\rightarrow 0$ rather than $S/T\rightarrow \infty$. 
The main issue is of course the claimed enormous disparity between 
the $(T)$ and $(S)$ contributions, not only the divergence of the 
$(S)$ contribution by itself, because both contributions must be 
small anyway, in order to be consistent with the observations.   

It was shown in detail in [10] (and, briefly, here) that 
the inflationary formula for density perturbations does not follow 
from correct dynamical equations plus correct quantum normalization. 
(This is one of the ``things everyone should know about inflation''.)
However, it is more or less clear without detailed calculations 
that the inflationary proposition can not be true.
The quantum-mechanical (parametric) generating mechanism can be described, 
both in quantum optics and in cosmology, as the depletion of a pump quantum 
into a pair of signal quanta. A weak laser can not generate an arbitrarily 
large number of squeezed photons at the expense of regulating a parameter 
unrelated to the strength of the pumping light. In contrast, 
the inflationary formula (33) suggests that a relatively weak gravitational 
field determined by a (fixed) $H^2$ can produce an arbitrarily large amount 
of density perturbations at the expense of regulating the denominator in
that formula, which does not change the strength of the pumping gravitational
field. The proposition expressed by formulas (33), (34) is known as the 
``standard inflationary result''. In the inflationary literature, 
ever since this claim was formulated on the grounds of dubious arguments, 
it is regularly being ``confirmed''. Big confusion has been added by
misinterpretation and abuse of the ``gauge-invariant'' formalism, which
resulted in the (mathematically incorrect and physically absurd) 
statement aimed at justification of the divergent ``standard result'', 
namely, that the long-wavelength scalar metric perturbations have 
experienced (in contrast to gravitational waves) an arbitrarily 
``big amplification'', proportional to $1/\sqrt\gamma$,
during a short ``reheating'' transition from 
$p =-\epsilon$ to $p=\epsilon /3$.

\section{Some Implications of Microwave Background Anisotropies
for Theory and Experiment}

The observed microwave background anisotropies signify the existence
in the Universe of small cosmological perturbations with extremely
long wavelengths. The wavelengths are of the order of, and longer than, 
the present-day Hubble radius $l_H$. It is argued~[2] 
that it is difficult to explain the origin of these perturbations 
without invoking the quantum-mechanical (parametric) mechanism 
for their generation. At any rate, in contrast to other possibilities, 
the quantum-mechanical (in fact, quantum gravitational) generation is, 
in a sense, an unavoidable process, since its theoretical foundation 
relies only on general relativity and basic principles of quantum 
field theory. This mechanism seems to be sufficient for explanation 
of the presently available data. Moreover, the comparison of theoretical 
predictions with the observations enables us to make certain further 
conclusions regarding the very early Universe and the expected outcomes 
of new forthcoming observations.

It is important to recall that the mere detection of the quadrupole 
anisotropy at the level $\delta T/T=5\times 10^{-6}$~[1] allows us
to conclude that, independently on the origin and nature of the 
responsible perturbations, the Universe remains to be homogeneous 
and isotropic at scales much larger than $l_H$ and up to distances 
about 500 times longer than $l_H$~[13]. At still longer scales, 
the homogeneity and isotropy of the Universe cannot be guaranteed, 
in the sense that some dimensionless deviations can be larger than 1 
without conflicting the CMB observations~[13]. 
The transition from spatially flat cosmological models to open models 
does not affect this conclusion considerably~[15].
Also, the mere existence of the long-wavelength cosmological perturbations
requires them to have special phases and to exist at the previous
radiation-dominated stage in the form of standing, rather than traveling,
waves~[2]. This conclusion follows from the Einstein equations, 
if we want to propagate the observed perturbations back in time up to, 
at least, the era of primordial nucleosynthesis without destroying 
the homogeneity and isotropy of that era. The distribution of phases 
can be only very narrow (highly squeezed) with two peaks separated 
by $\pi$. This distribution of phases arises inevitably during 
the generation of squeezed vacuum states, and is known in formal 
quantum mechanics and quantum optics as the ``phase bifurcation''~[16]. 

The processing of the COBE data has resulted in evaluation of the
power-law spectral index of the long-wavelength perturbations. 
In the context of density perturbations, the spectral index is often 
denoted by $n$ (not to be confused with the wavenumber $n$), 
and the Harrison-Zeldovich-Peebles spectrum (or ``flat'' spectrum) 
corresponds to $n=1$. This convention is caused by the description 
of the variance for $\delta \rho /\rho$
in terms of the wavelength intervals, not in terms of the logarithmic
wavelength intervals, used above. The transition to the characteristic
metric amplitudes and logarithmic intervals would make the 
Harrison-Zeldovich-Peebles spectral index equal to zero, 
in agreement with the ``flat'' nature of that spectrum. 
In any case, the exact relationship between this spectral index $n$ 
and the parameter $\beta$ used above is
\begin{equation}
   n \equiv 2\beta + 5 \quad ,
\end{equation}
so that $n=1$ corresponds to $\beta=-2$. The authors of Ref.~[17]
derived from the COBE data
\begin{equation}
  n = 1.2\pm 0.3\quad ,
\end{equation}
whereas the authors of~[18] derived from the same data, but processed 
in a different manner,
\begin{equation}
 n = 1.84\pm 0.29\quad .
\end{equation}

In these evaluations, we see the indication that the true value
of the spectral index $n$ is larger than 1, $n>1$. 
Assuming that the relevant cosmological perturbations were generated 
quantum-mechanically, the $n>1$ translates into $\beta > -2$, see Eq.~(36). 
Specifically, the value $n=1.4$, intermediate between the estimates (37) 
and (38), requires $\beta =-1.8$ and $p=-1.2\epsilon$ (see Eq.~(26)), 
whereas the conservative value $n=1.2$ requires $\beta =-1.9$ and 
$p= -1.1\epsilon$.  With the effective equations of state of this kind, 
$p+\epsilon < 0$, the initial stage of expansion was accompanied 
by growth of energy density and curvature of the space-time. 
It seems to the author that the search for a ``microphysical'' 
model of primordial matter capable of producing this kind of effective 
equations of state becomes an important theoretical problem.
This primordial matter should also be capable of supporting 
the scalar-type metric oscillations properly coupled to the pumping 
gravitational field, if we wish to apply the quantum-mechanical 
generating mechanism for the production of density perturbations. 
It is likely that the basic dynamical equation (20) will still be valid, 
with the replacement of $\gamma$ by $|\gamma|$.

Certainly, the required equation of state cannot be accommodated
by a scalar field, which is the basis of inflationary models.
If the spectral index $n>1$ is confirmed, this will mean that the
very early Universe was not driven by a scalar field. This conclusion
is true irrespective of the form of the scalar field potential 
$V(\phi )$.  Indeed, the energy density and pressure produced 
by any scalar filed are given by  
\[
  \epsilon = {1\over 2c^2} \dot{\phi}^2 + V(\phi )~~~,\quad
         p = {1\over 2c^2} \dot{\phi}^2 - V(\phi )\quad .
\]
This means that 
$\epsilon +p={1\over c^2}\dot{\phi}^2 \geq 0$, in contrast to the
required $\epsilon +p < 0$.

The experimentally evaluated spectral index $n$ and, hence, the
evaluated parameter $\beta$ enable us to make a relatively firm prediction
with regard to expected amplitudes of relic (squeezed) gravitational waves 
in higher frequency intervals. This prediction is based on the assumptions
that the perturbations responsible  for $\delta T/T$ were generated 
quantum-mechanically, and that the one and the same cosmological model,
determined by the derived fixed $\beta$, is responsible for the entire 
spectrum, Eqs.~(27)--(29). Taking into account all known theoretical and 
observational arguments, it is difficult to avoid these input assumptions,
but, strictly speaking, they are, so far, assumptions. Since the
contribution of gravitational waves to the large-angular-scale anisotropies
is not smaller (in fact, it is somewhat bigger) than the contribution of
density perturbations, and since their spectral indices are strictly
the same (Eqs.~(27), (30)), the measured anisotropies can be taken as
experimental data about gravitational waves. The evaluated parameter
$\beta$ can now be used in formula (29) for $h(\nu )$ in frequency intervals 
accessible for ground-based and space laser interferometers~[2].

We will start from $n=1.4$ and $\beta =-1.8$. (For some theoretical 
considerations about a possibility for $n$ to be close to $n=1.4$, 
see~[19].)  In this case, the predicted signal is well above 
the expected sensitivity of the proposed space interferometer LISA~[20] 
and the advanced ground-based LIGO (see~[3] and references there). 
Indeed, the expected signal is 
$h(\nu )=10^{-19}$, $\Omega_g(\nu )=10^{-8}$ 
at the LISA-tested frequency
$\nu =10^{-3}$~Hz, and $h(\nu )=10^{-23}$, $\Omega_g(\nu )=10^{-6}$ 
at the LIGO/VIRGO/GEO-tested frequency $\nu = 10^2$~Hz. 
If the more conservative value 
$n=1.2$, $\beta =-1.9$ is confirmed, 
this will still provide a measurable signal. Namely, 
$h(\nu ) =10^{-20.5}$, $\Omega_g(\nu ) = 10^{-11}$ at
$\nu = 10^{-3}$~Hz and $h(\nu )=10^{-25}$, $\Omega_g(\nu )=10^{-10}$
at $\nu = 10^2$~Hz.

Thus, in the framework of the theory of quantum-mechanically
generated cosmological perturbations, gravitational waves are largely    
responsible for the measured large-angular-scale anisotropies, and a
more accurate determination (within certain limits~[21]) of the spectral
index will make it possible to give more accurate estimates for the
higher-frequency relic gravitational waves.

It is necessary to say that the ``Pre-Big-Bang'' cosmological
scenario~[22], based on a non-Einstein gravitational theory, does also
operate with the scale factors having $\beta > -2$. However, according
to calculations of the authors of this scenario~[22], it appears that
the quantum-mechanically generated gravitational waves exhibit a peak 
at high frequencies, $\nu =(10^5-10^{10})$~Hz. The calculated spectrum 
is far too weak at lower frequencies, and is unable to affect and explain
the large-angular-scale anisotropies~[22].

A convincing test of the truly quantum-mechanical origin of certain 
cosmological perturbations should probably exploit the most distinct 
properties of the generated squeezed vacuum quantum states:
small variances of phase and large variances of amplitude, and their
statistical properties in general. The small variances of phase are
reflected in the standing-wave pattern of the generated fields. 
This is related to such phenomena as the Sakharov oscillations 
in the density perturbation spectrum and associated oscillations 
of the higher index multipole components of the CMB anisotropies. 
This property is also responsible for the nonstationary character 
of the relic high-frequency gravitational wave noise. etc. 
The large variances of amplitude are reflected in statistical 
distributions of various measurable quantities,
such as the angular correlation variable \\
$v={\delta T\over T}({\bf e}_1){\delta T\over T}({\bf e}_2)$, etc. 

It appears that we are enforced to deal with the quantum gravity
processes, and a detailed comparison of their theoretical predictions
with observations will certainly be a fascinating area of research in
the coming years.
\\

\leftline{\bf Acknowledgments}

I appreciate the hospitality of Professor J.~Helayel-Neto and the 
Department of Fields and Particles at CBPF in Brazil, 
where a part of this paper was written. 
The final preparation of the manuscript was facilitated by the
skillful assistance of Pranoat Suntharothok-Priesmeyer.
\\

\leftline{\bf References}
\begin{itemize}
\item[1.] 
G. F. Smoot {\it et al.,} Astrophys. J. {\bf 396}, L1 (1992);\\
C. L. Bennet {\it et al.,} Astrophys. J. {\bf 436}, 423 (1994);\\
E. L. Wright {\it et al.,} Astrophys. J. {\bf 436}, 443 (1994).
\item[2.]
L. P. Grishchuk, Class. Quant. Gravity {\bf 14}, 1445 (1997).
\item[3.]
K. S. Thorne, Report gr-qc/9706079.
\item[4.]
R. Loudon and P. Knight, 
J. Modern Opt. {\bf 34}, 709 (1987);\\
P. Knight, in {\it Quantum Fluctuations}, Eds. E.~Giacobino, S.~Reynauld,
and J.~Zinn-Justin (Elsevier Science, 1997).
\item[5.]
L. Grishchuk, H. A. Haus, and K. Bergman, 
Phys. Rev. {\bf D 46}, 1440 (1992); \\
L. P. Grishchuk, in {\it Quantum Fluctuations}, 
Eds. E. Giacobino, S. Reynauld, and J. Zinn-Justin 
(Elsevier Science, 1997).
\item[6.] 
L. D. Landau and E. M. Lifshitz, {\it The Classical Theory of Fields}
(Pergamon, New York, 1975).
\item[7.] 
L. P. Grishchuk, Lett. Nuovo Cimento {\bf 12}, 60 (1975);
Zh. Eksp. Teor. Fiz. {\bf 67}, 825 (1974) 
[Sov. Phys. JETP {\bf 40}, 409 (1975)];
Ann. NY Acad. Sci. {\bf 302}, 439 (1977).
\item[8.]
S. L. Shapiro and S. A. Teukolsky, 
{\it Black Holes, White Dwarfs and Neutron Stars} (Wiley, New York, 1983).
\item[9.] 
L. P. Grishchuk, in {\it Current Topics in Astrofundamental Physics}, 
Eds. N. Sanchez and A. Zichichi (World Scientific, Singapore, 1992), 
p.~435.
\item[10.] 
L. P. Grishchuk, Phys. Rev. {\bf D 50}, 7154 (1994); 
in {\it String Gravity and Physics at the Planck Energy Scale}, 
Eds. N. Sanchez and A. Zichichi (Kluwer, Dordrecht, 1996), p.~369;
Phys. Rev. {\bf D 53}, 6784 (1996).
\item[11.] 
S. W. Hawking, Phys. Lett. {\bf 115B}, 295 (1982);\\
A. A. Starobinsky, Phys. Lett. {\bf 117B}, 175 (1982);\\
A. Guth and S.-J. Pi, Phys. Rev. Lett. {\bf 49}, 1110 (1982);\\
J. M. Bardeen, P. Steinhardt, and M. Turner, Phys. Rev. {\bf D 28},
679 (1983).
\item[12.]
R. K. Sachs and A. M. Wolfe, Astrophys. J. {\bf 1}, 73 (1967).
\item[13.]
L. P. Grishchuk and Ya. B. Zeldovich, Astron. Zh. {\bf 55}, 209 (1978)
[Sov. Astron, 22, 125 (1978)];\\
L. P. Grishchuk, Phys. Rev. {\bf D 45}, 4717 (1992).
\item[14.] 
V. A. Rubakov, M. V. Sazhin, and A. V. Veryaskin, Phys. Lett. {\bf 115B},
189 (1982);\\ 
R. Fabbri and M. D. Pollock, Phys. Lett. {\bf 125B}, 445 (1983).
\item[15.] 
J. Garcia-Bellido, A. R. Liddle, D. H. Lyth, and D. Wands, 
Phys. Rev. {\bf D 52}, 6750 (1995).
\item[16.] 
W. Schleich and J. A. Wheeler, J. Opt. Soc. Am. {\bf B 4}, 1715 (1987);\\
W. Schleich, R. J. Horowicz, and S. Varro, 
Phys. Rev. {\bf A 40}, 7405 (1989).
\item[17.] 
C. L. Bennett {\it et al.,} Astrophys. J. {\bf 464}, L1 (1996);\\
K. M. Gorski {\it et al.,} Astrophys. J. {\bf 464}, L11 (1996);\\
G. Hinshaw {\it et al.,} Astrophys. J. {\bf 464}, L17 (1996).
\item[18.]
A. A. Brukhanov {\it et al.,} Report astro-ph/9512151.
\item[19.] 
I. Antoniadis, P. O. Mazur, and E. Mottola, Phys. Rev. Lett. {\bf 79},
14 (1997).
\item[20.] 
P. Bender {\it et al.} LISA Pre-Phase A Report, MPG 208, 1996.
\item[21.]
L. P. Grishchuk and J. Martin, Phys. Rev. {\bf D 56}, 1924 (1997).
\item[22.] 
M. Gasperini and G. Veneziano, Astropart. Phys. {\bf 1}, 317 (1993);\\
M. Gasperini and M. Giovannini, Phys. Rev. {\bf D 47}, 1519 (1993);\\
G. Veneziano, in {\it String Gravity and Physics at the Planck Energy 
Scale}, Eds. N. Sanchez and A. Zichichi (Kluwer, Dordrecht, 1996), p.~285;\\
M. Gasperini, Report gr-qc/9707034.
\end{itemize}
\end{document}